\documentclass[review]{elsarticle}

\biboptions{authoryear}

\usepackage{url}
\usepackage{hyperref}
\usepackage{algorithmic}
\usepackage{algorithm}
\usepackage{theorem}
\usepackage{amsmath,amssymb}
\usepackage{graphicx}
\usepackage[misc,geometry]{ifsym}
\usepackage{color}
\usepackage{array}

\newtheorem{definition}{Definition}
\newtheorem{theorem}{Theorem}
\newtheorem{proposition}{Proposition}
\newtheorem{example}{Example}

\newtheorem{corollary}{Corollary}
\newtheorem{problem}{Problem}

\newcommand{\BV}{\underline{{\mathfrak B}}}
\newcommand{\K}{\mathbb{K}}
\newcommand{\R}{\mathcal{R}}
\newcommand{\D}{\mathbb{D}}
\newcommand{\F}{\mathcal{F}}

\newcommand{\Proof}[1]{\noindent \textbf{Proof.}{#1}{$\square$}}

\definecolor{obj-red}{rgb}{1,0,0}
\definecolor{obj-blue}{rgb}{0,0,1}
\definecolor{obj-green}{rgb}{0,0.5,0}

\journal{Discrete Applied Mathematics}

\begin{document}

\begin{frontmatter}

\title{On closure operators related to maximal tricliques in tripartite hypergraphs} 

\author{Dmitry I. Ignatov}

\begin{abstract}

Triadic Formal Concept Analysis (3FCA) was introduced by Lehman and Wille almost two decades ago. And many  researchers work in Data Mining and Formal Concept Analysis using the notions of closed sets, Galois and closure operators, closure systems, but up-to-date even though that different researchers actively work on mining triadic and n-ary relations, a proper closure operator for enumeration of triconcepts, i.e. maximal triadic cliques of tripartite hypergaphs, was not introduced. In this paper we show that the previously introduced operators for obtaining triconcepts and maximal connected and complete sets (MCCSs) are not always consistent and provide the reader with a definition of valid closure operator and associated set system. Moreover, we study the difficulties of related problems from order-theoretic and combinatorial point view as well as provide the reader with justifications of the complexity classes of these problems.
  
\end{abstract}


\end{frontmatter}


\section{Introduction}
Pattern mining is one of the most important Data Mining areas and often relies on fundamental notions from theoretical computer science and algebra like fixpoints, closure operators and lattices \citep{Zaki:2005a,Boley:2010}. Formal Concept Analysis \citep{Ganter:1999} can be considered as an elegant algebraic framework to deal with (frequent) closed sets of objects and their attributes (formal concepts or maximal bicliques) by means of two closure operators formed by Galois connection over these sets. 

Recent studies showed that there are efficient algorithms for building all formal concepts not only in binary object-attribute case but in ternary (\textsc{TRIAS}, \citep{Jaschke:2006}) and $n$-ary cases (\textsc{Data-Peeler}, \citep{Cerf:2009}). 

Several researchers tried to develop a proper closure operator for triadic \citep{Trabelsi:2012} and n-ary cases \citep{Spyropoulou:2014}. However the detailed analysis in this paper shows that the concept-forming operator in \citep{Trabelsi:2012} is not always monotone on triset systems. An interesting approach from \citep{Spyropoulou:2014} can be used to enumerate formal triconcepts as the maximal fixpoints of a set system of closed and connected sets (CCS) but suffers from presence of phantom hyperedges because of the lossy $k$-partite graph encoding. In this paper, we show how to define a proper triset system for the concept forming operator from \citep{Trabelsi:2012} that makes it a closure operator, describe the family of closure operators of this type and investigate their properties, and prove that there is no an associated closure operator on the whole triset system for a given tricontext. We also introduce a notion of (maximal) switching generator -- a triset resulting in different closed patterns that contain it. In addition we show how to deal with lossy hyperedge encoding and phantom edges to generate triconcepts as maximal connected and complete sets.

The rest of the paper is organised as follows. In Section~\ref{sec:MMP}, we recall basic definitions from FCA and its polyadic extensions and reproduce necessary definitions and propositions from \citep{Spyropoulou:2014}. In Section~\ref{sec:recclos}, we discuss the studied concept and closed CCS forming operators with a focus on their inconsistency conditions. Section \ref{sec:triclo} reports our main results. Section~\ref{sec:relw} discusses related work and and Section~\ref{sec:concl} concludes the paper.

\section{Multimodal and multirelational closed patterns}
\label{sec:MMP}

\subsection{Formal Concept Analysis and its polyadic extensions}

First, we  recall some basic notions from Formal Concept Analysis
(FCA) \citep{Ganter:1999}.

 Let $G$ and $M$ be sets, called the
set  of objects and attributes, respectively, and let $I$ be a
relation $I\subseteq G\times M$: for $g\in G, \ m\in M$,  $gIm$
holds  iff the  object  $g$  has  the attribute $m$. The triple
$\mathbb{K}=(G,M,I)$ is called a \emph{(formal) context}.

A \emph{triadic context} $\mathbb{K}=(G,M,B,Y)$ consists of sets $G$ (objects), $M$ (attributes), and $B$ (conditions), and ternary relation $Y\subseteq G \times M \times B$ \citep{Lehmann:1995}. An incidence $(g, m, b) \in Y$ shows that the object $g$ has the attribute $m$ under the condition $b$.

An \emph{$n$-adic context} is an $(n + 1)$-tuple $\K = (K_1,K_2, \ldots,K_n, Y)$, where $Y$ is an $n$-ary relation between sets $K_1, \ldots , K_n$ \citep{Voutsadakis:02}.

\subsubsection{Concept forming operators and formal concepts}
\label{sec:1b}

If $A\subseteq G$,
$B\subseteq  M$ are arbitrary subsets of objects and attributes, respectively, then the {\it Galois
connection} is given by the following {\it derivation operators}:

\begin{eqnarray}
\begin{array}{c}
A' = \{m\in M\mid  gIm \ {\rm for\ all}\ g\in A\}, \\
B' = \{g\in G\mid  gIm \ {\rm for\ all}\ m\in B\}.
\end{array}
\end{eqnarray}

If we have several contexts, the derivation operator of a context $(G, M, I)$ is denoted by $(.)^I$.

The pair $(A,B)$, where $A\subseteq G$, $B\subseteq M$, $A' = B$,
and $B' = A$ is called a {\it (formal) concept (of the context
$\mathbb{K}$)} with {\it extent} $A$ and {\it intent} $B$ (in this case we
have also $A'' = A$ and $B'' = B$).

The concepts, ordered by $(A_1,B_1)\geq (A_2,B_2) \iff A_1\supseteq A_2 (B_2\supseteq B_1)$, form a complete  lattice, called \emph{the concept lattice}  $\BV(G,M,I)$.

\subsubsection{Formal concepts in triadic and in n-ary contexts}
\label{sec:1c}

For convenience, a triadic context is denoted by $\mathbb{K}=(X_1,X_2,X_3,Y)$\footnote{Note that in the title we refer to a formal tricontext as a tripartite hypergraph since we deal with three types of vertices connected by triadic hyperedges.}. A triadic context $\mathbb{K}=(X_1,X_2,X_3,Y)$ gives rise to the following dyadic contexts

$\mathbb{K}^{(1)}=(X_1, X_2\times X_3, Y^{(1)})$,
$\mathbb{K}^{(2)}=(X_2, X_1\times X_3, Y^{(2)})$,
$\mathbb{K}^{(3)}=(X_3, X_1\times X_2, Y^{(3)})$,

where $gY^{(1)}(m,b):\Leftrightarrow mY^{(2)}(g,b):\Leftrightarrow bY^{(3)}(g,m):\Leftrightarrow (g,m,b) \in Y$. The derivation operators (primes or concept-forming operators) induced by $\mathbb{K}^{(i)}$ are denoted by $(.)^{(i)}$. For each induced dyadic context we have two kinds of such derivation operators. That is, for $\{i,j,k\}=\{1,2,3\}$ with $j<k$ and for $Z \subseteq X_i$ and $W \subseteq X_j\times X_k$, the $(i)$-derivation operators are defined by:

$$Z \mapsto Z^{(i)} = \{(x_j,x_k) \in X_j\times X_k| x_i, x_j, x_k \mbox{ are related by Y for all } x_i \in Z\},$$
$$W \mapsto W^{(i)} = \{x_i \in X_i| x_i, x_j, x_k \mbox{ are related by Y for all } (x_j,x_k) \in W\}.$$
Formally, a triadic concept of a triadic context $\mathbb{K}=(X_1,X_2,X_3,Y)$ is a triple $(A_1,A_2,A_3)$ of $A_1 \subseteq X_1, A_2 \subseteq X_2, A_3 \subseteq X_3$ such that for every $\{i,j,k\}=\{1,2,3\}$ with $j<k$ we have $(A_j \times A_k)^{(i)}=A_i$.
For a certain triadic concept $(A_1,A_2,A_3)$, the components $A_1$, $A_2$, and $A_3$ are called the extent, the intent, and the modus of $(A_1,A_2,A_3)$. It is important to note that for interpretation of $\mathbb{K}=(X_1,X_2,X_3,Y)$ as a three-dimensional cross table, according to our definition, under suitable permutations of rows, columns, and layers of the cross table, the triadic concept $(A_1,A_2,A_3)$ is interpreted as a maximal cuboid full of crosses.

 The set of all triadic concepts of $\mathbb{K}=(X_1,X_2,X_3,Y)$ is denoted by $\mathfrak{T}(\mathbb{K})$. However this set does not form a partial order by extent inclusion since it is possible for the same triconcept extent to have different combinations of intent and modus components \citep{Wille:1995,Lehmann:1995}, similarly, for orderings along the attribute and condition components.

There is a quasiorder $\lesssim_i$  for each $i \in \{1, 2, 3\}$ and its corresponding equivalence relation $\sim_i$ is defined by
$$(A_1,A_2,A_3) \lesssim_i  (B_1,B_2,B_3): \Longleftrightarrow  A_i \subseteq B_i \mbox{ and } $$
$$(A_1, A_2, A_3) \sim_i (B_1, B_2, B_3): \Longleftrightarrow A_i = B_i.$$

These quasiorders satisfy the antiordinal dependencies \citep{Wille:1995}: For $\{i,j,k\} = \{1,2,3\}$ and all triconcepts $(A_1,A_2,A_3)$ and $(B_1,B_2,B_3)$ from $\mathfrak{T}(\mathbb{K})$ it holds that $(A_1, A_2, A_3)  \lesssim_i (B_1, B_2, B_3)$ and $(A_1, A_2, A_3)  \lesssim_j (B_1, B_2, B_3)$ imply $(A_1, A_2, A_3)  \gtrsim_k (B_1, B_2, B_3)$.

One may introduce $n$-adic formal concepts without $n$-ary concept forming operators. The $n$-adic concepts of an $n$-adic context $(K_1, \ldots ,K_n, Y)$ are exactly the maximal $n$-tuples $(A_1, \ldots , A_n)$ in $2^{K_1} \times \cdots \times 2^{K_n}$, where $A_1 \times \cdots \times A_n \subseteq Y$ with respect to component-wise set inclusion \citep{Voutsadakis:02}. The notion of $n$-adic concept lattice can be introduced similarly to the triadic case \citep{Voutsadakis:02}.

\subsection{Maximal closed connected sets}
Here we introduce necessary defintions and results from a series of papers on mining maximal closed connected sets \citep{Spyropoulou:2014,Lijffijt:2016}. Note that the authors prefer to use terminology close to relational databases but the main definitions can be easily reproduced in terms of $k$-partite graphs; to find related works in FCA community one may refer to Relational Concept Analysis \cite{Hacene:2013}.

\cite{Spyropoulou:2014} formalised a multi-relational database (MRD) as a tuple $\D =
(E, t, \mathcal{R}, R)$, where $E$ is a finite set of entities that is partitioned into $n$ entity types by a
mapping $t: E \to \{1, \ldots, n\}$, i.e., $E = E_1 \sqcup
\cdots \sqcup E_k$\footnote{Here and later, $\sqcup$ means disjoint union.} with $E_i = \{e \in E\mid t (e) = i \}$.
Moreover, $R \subseteq \{\{i, j\}\mid i, j \in \{1, \ldots , k\}, i \neq j \}$ is a set of relationship types such
that for each $\{i, j\} \in R$ there is a binary relation $\mathcal{R}_{\{i, j\}} \subseteq \{\{e_i , e_j\}\mid e_i \in E_i , e_j \in
E_j \}$. The set $\mathcal R$ then is the union of all these relations, i.e., $\mathcal{R} = \bigcup_{\{i, j\} \in R}
\R_{\{i, j \}}$.
This definition allows relationship types can be many-to-many, one-to-many, or one-to-one, depending on
how many relationships the entities of either entity types can participate in. The authors do not allow relationship types between an entity type and itself since they mainly concentrate on relations between entities of different types, but the former can be modeled by having two copies of the same entity type
and a relationship type between them.

\begin{definition} (Completeness) \citep{Spyropoulou:2014} A set $F \subseteq E$ is complete if for all $e, \tilde{e} \in F$ with
$\{t(e), t(\tilde{e})\} \in R$ it holds that $\{e, \tilde{e}\} \subseteq \mathcal{R}_{\{t (e),t (\tilde{e}
)\}} $.
\end{definition}

\begin{definition} (Connectedness) \citep{Spyropoulou:2014} A set $F \subseteq E$ is connected if for all $e, \tilde{e} \in F$ there is a
sequence $e = e_1, \ldots , e_l = \tilde{e}$ with $\{e_1, \ldots , e_l\} \subseteq F$ such that for $i \in \{1, \ldots , l \}$
it holds that $\{e_i , e_{i+1}\} \in R$.
\end{definition}

It implies that a subset of size larger than one can be connected only if it contains entities
of at least two different types.

A set $F \subseteq E$ is a Complete Connected Subset (CCS) if it satisfies both connectedness
and completeness.

A Maximal Complete Connected Subset (MCCS) is a CCS to which no
element can be added without violating connectedness or completeness.

For a database $\D = (E,t,\R, R)$
the set system of CCSs, is defined as
$\mathcal{F}_\D = \{F \subseteq E\mid F \mbox{ is connected and complete} \}$.
From an algorithmic point of view, the property of strong accessibility means that for two CCSs $X, Y \in \mathcal{F}_\D$
with $X \subseteq Y$ , it is possible to iteratively extend $X$ by one element at a time, only
passing via sets from the set system and finally obtain $Y$. Formally, for a set
system $F \subseteq 2^A$, where $A$ is the ground set, and a set $F \in \F$, let us denote by
$Aug(F) = \{a \in A\mid F \cup \{a\} \in \F\}$ the set of valid augmentation elements of $F$.
Then $\F$ is called strongly accessible  if for all $X \subset Y \subseteq A$ with $X, Y \in \F$ there is
an element $e \in (Aug(X) \setminus X) \cap Y$. 

\begin{theorem} \citep{Spyropoulou:2014} For all relational databases $\D = (E, t,\mathcal{R}, R)$, the set system $\mathcal{F}_\D$ of CCSs
is strongly accessible.
\end{theorem}

Specifically for the set system $\mathcal{F}_\D$ of CCSs, and given
a relational database $\D = (E, t,\R, R)$, the set $Aug(F)$ corresponds to the following
set: $Aug(F) = \{e \in E\mid F \cup \{e\} \mbox{ is complete and connected} \}$.
Note that for the sake of efficiency $Aug(F)$ can be recursively updated.

To define a closure operator for the set system $\F_\D$ the authors 
make use of the set of compatible entities which is defined as follows:

\begin{definition} (Compatible entities) \citep{Spyropoulou:2014} For a relational database $\D = (E, t, \R, R)$ the
set of compatible entities of a set $F \in \F_\D$ is defined as $Comp(F) = \{e \in E \mid
F \cup {e} \mbox{ is complete}\}$.
\end{definition}

\begin{definition} ($g$ operator) \citep{Spyropoulou:2014} For a relational database $\D = (E, t,\R, R)$ the
operator $g : \F_\D \to 2^E$ is defined as
$g(F) = \{ e \in Aug(F)\mid Comp(F \cup {e}) = Comp(F) \}$.
\end{definition}

\begin{proposition} \citep{Spyropoulou:2014} For all relational databases $\D = (E, t, \R, R)$, the codomain of the $g$
operator is the set system $\F_\D$ of CCSs and $g$ is extensive and monotone.
\end{proposition}

\begin{proposition} \label{prop:idemp}  \citep{Spyropoulou:2014} For all relational databases $\D = (E, t, \R, R)$ with the property that $e \in E$ such that $\{e\} \cup E_i$ is complete and connected for an $i \in t(E)$, the operator $g$ is idempotent.
\end{proposition}

\begin{corollary} \citep{Spyropoulou:2014} For all relational databases $\D = (E, t,\R, R)$, with the property that
$e \in E$ such that $\{e\} \cup E_i$ is complete and connected for an $i \in t(E)$, the operator $g$
is a closure operator.
\end{corollary}

Note that, the technical requirement in Proposition~\ref{prop:idemp} for $g$ being idempotent may be fulfilled by adding an isolated vertex $\{e_0\}$ to $E_i$  for all $E_i \subseteq E$ and $e \in E\setminus E_i$, where $E_i \cup \{e\}$ is CCS.

\begin{figure}[h]
	\centering
		\includegraphics[width=\textwidth]{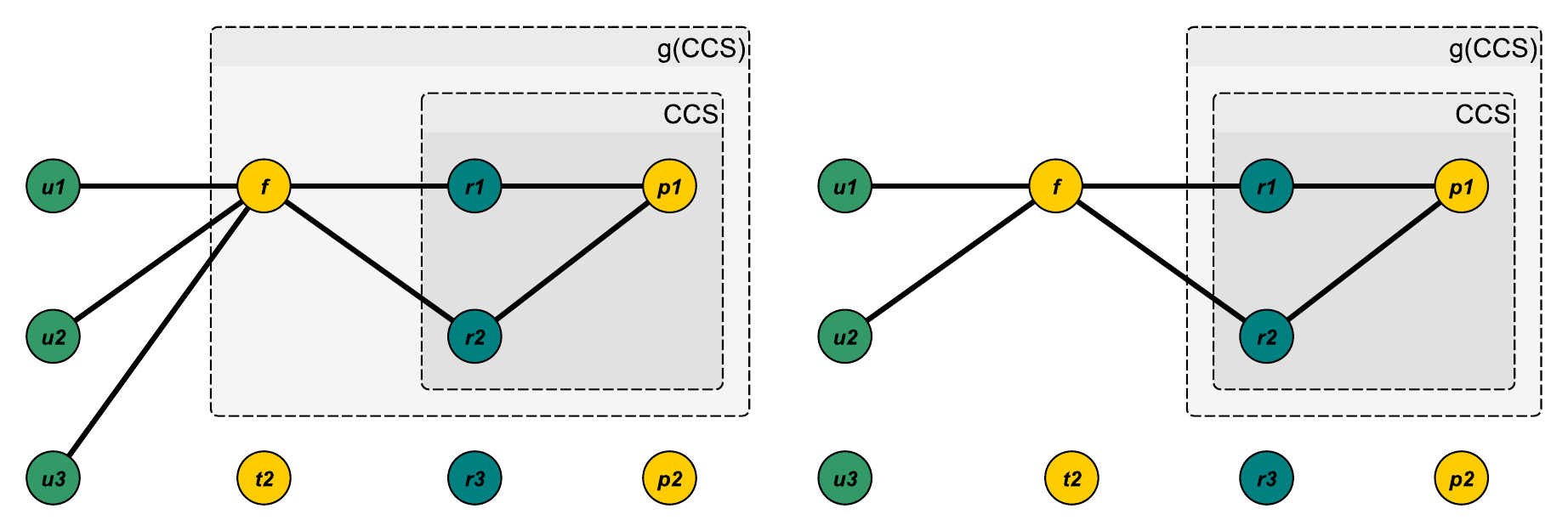}
	\caption{On idempotency of $g(\cdot)$}
	\label{fig:Idempotency}
\end{figure}

\begin{example}
In Figure~\ref{fig:Idempotency}, on the left one can see the  violation of idempotency of $g(\cdot)$ since $g(\{r_1,r_2,p_1\})=\{r_1,r_2,p_1,f\}$ and 
 $g(g(\{r_1,r_2,p_1\}))=\{r_1,r_2,p_1,f, u_1, u_2, u_3\}$. On the right graph of Figure \ref{fig:Idempotency} the idempotency fulfills  since $g(\{r_1,r_2,p_1\})=g(g(\{r_1,r_2,p_1\}))$ $=\{r_1,r_2,p_1\}$. It happens since for the left graph $$Comp(\{r_1,r_2,p_1\}\cup f)=\{r_1,r_2,p_1,f, u_1, u_2, u_3\}=Comp(\{r_1,r_2,p_1\}),$$ but for the right one $$Comp(\{r_1,r_2,p_1\}\cup f)=\{r_1,r_2,p_1,f, u_1, u_2\} \neq Comp(\{r_1,r_2,p_1\})=\{r_1,r_2,p_1,f, u_1, u_2, u_3\}.$$
\end{example}


\section{Pitfalls of recent candidates for closure operators in triadic case}\label{sec:recclos}

\subsection{Non-monotonicity of TriCons concept forming operator}

To simplify further considerations of tri-sets, triadic concepts and multirelational databases both as tuples and sets, we introduce two interrelated operators.

\begin{definition} \citep{Trabelsi:2012} Let $\K=(G,M,B,I)$ be a formal tricontext. A triple  $(X,Y,Z)$   is called a triset of $\K$ iff $ X \times Y \times Z \subseteq I$.
\end{definition}

Note that \cite{Cerf:2009} define a triset of $\K$ differently: $ X \times Y \times Z \in 2^G \times 2^M \times 2^B$. We keep the former definition to work with $h(\cdot)$ in the original setting \citep{Trabelsi:2012}.  

Note that according the definition of Cartesian product, if at least one of the sets $X,Y$ or $Z$ is $\emptyset$ \citep{Simovici:2008}, then $X \times Y \times Z = \emptyset$, so $\emptyset \subseteq I$ and $(X,Y,Z)$ is a triset. However trisets $(X,Y,\emptyset)$ and $(X, \emptyset, Z)$ have different structure even though $X \times Y \times \emptyset = X \times \emptyset \times Z = \emptyset \subseteq I$.

\begin{definition} For a formal tricontext $\K=(G,M,B,I)$ and any triple $(X,Y,Z) \subseteq 2^G\times 2^M \times 2^B$ (e.g. triconcept) of $\K$ the operator $flat: 2^G\times 2^M \times 2^B \to 2^{G\sqcup M \sqcup B}$ is defined as follows: $flat(X,Y,Z)= X \sqcup Y \sqcup Z$. 
\end{definition}

\begin{definition} For a given formal n-context $\K=(E_1,\ldots,E_n, I \subseteq E_1 \times \ldots \times E_n)$ (or multi-relational database $\D=(E,t,\R,R)$), where $E=flat(E_1,\ldots,E_n)$, and $S \subseteq 2^{E}$,  the operator $tuple: 2^E \to 2^{E_1} \times \ldots \times 2^{E_n}$ is defined as follows: $tuple(S)= (E_1 \cap S, \ldots, E_n \cap S)$. 
\end{definition}

Triple compositions of $tuple(\cdot)$ and $flat(\cdot)$ operators form identity operators $tuple(flat(tuple(\cdot)))=id_S(\cdot)$ and $flat(tuple(flat()))=id_T(\cdot)$ over sets and tuples respectively.

Note that for trisets $t_1=(A_1,B_1,C_1)$ and  $t_2=(A_2,B_2,C_2)$, $t_1 \sqsubseteq t_2$  means that $A_1 \times B_1 \times C_1   \subseteq A_2 \times B_2 \times C_2$, i.e. every triple  $(a,b,c)  \in (A_1,B_1,C_1)$  is in $(A_2,B_2,C_2)$. It follows that $\sqsubseteq$ is not antisymmetric, since e.g. $(X,Y, \emptyset) \sqsubseteq (X,\emptyset,Z)$ and $(X,\emptyset,Z) \sqsubseteq (X,Y, \emptyset)$, but $(X,Y, \emptyset)\neq (X,\emptyset,Z)$. Thus every preorder $(\mathcal T \subseteq 2^G \times 2^M \times 2^B \cap 2^I, \sqsubseteq)$ have all equivalence classes of cardinality 1 except $[\emptyset]=\{(\emptyset, \emptyset, \emptyset ), \ldots , (G,\emptyset,\emptyset) \ldots, (\emptyset,M,B)\}$ of cardinality $2^{|G|+|M|+|B|}-1$.

\begin{definition} \citep{Trabelsi:2012} Let $S=(X, Y, Z)$ be a tri-set of $\K=(G,M,B, I \subseteq G \times M \times B)$. The mapping $h: 2^G\times 2^M \times 2^B \cap  2^I \to 2^G\times 2^M \times 2^B$ is defined as follows:

$h(S) = \{(U, V , W) \mid  U = \{g \in G \mid  \forall m \in Y, \forall b \in Z: (g, m, b) \in I \}$
$\wedge V = \{m \in M \mid  \forall g \in U, \forall b \in Z: (g, m, b) \in Y \}$
$\wedge W = \{b \in B \mid  \forall g \in U, \forall m \in V: (g, m, b) \in Y \}$

\end{definition}

Note that every triconcept is a maximal or closed triset, i.e. a triset that cannot be extended by triples from $I$ being a triset.

\begin{proposition} $h(\cdot)$ is extensive and idempotent by $\sqsubseteq$ on $ T =\{t \mid   t \mbox{ is a triset of } \K \}=\{ (X,Y,Z) \in 2^G\times 2^M \times 2^B \mid  (X,Y,Z) \subseteq I \}$ and every fixpoint $f$ of $h$ (i.e. $h(f)=f$) is a triconcept of $\K$. 

\end{proposition}
\Proof{
One can find the proof of extensivity and idempotency in \citep{Trabelsi:2012}. It is easy to see that every formal triconcept is a fixpoint of $h(\cdot)$ and every triset $(X,Y,Z)$ is transformed by $h(\cdot)$ to the triconcept $((Y\times Z)^{(1)},((Y\times Z)^{(1)} \times Z)^{(2)},((Y\times Z)^{(1)} \times ((Y\times Z)^{(1)} \times Z)^{(2)})^{(3)}$. Indeed,  all formal triconcepts should be listed since  a triset is allowed to be a triple with at least one component being $\emptyset$.
}

\begin{theorem} For a given tricontext  $\K=(G,M,B, I \subseteq G \times M \times B)$ and its associated triset system $\mathcal T=\{ (X,Y,Z)   \in 2^G\times 2^M\times 2^B \mid  (X,Y,Z) \subseteq I \}$ operator $h$ is not monotone w.r.t. $\sqsubseteq$. 

\end{theorem}
\textbf{Proof.} To construct a violating example, one needs two different triconcepts with the same extent, $c_1=(X,Y_1,Z_1)$ and $c_2=(X,Y_2,Z_2)$ of $\K$ such that $Y_1 \subset Y_2$ and $Z_1 \supset Z_2$.

Consider the tri-set $s=(X, Y_1, Z_2)$: 
$$s \sqsubseteq c_1 \Rightarrow h(s)=c_2 \not\sqsubseteq h(c_1)=c_1$$ $\square$.

\begin{example}

For the tricontext in Figure~\ref{counterEx1}, the violating example for monotonicity of $h(\cdot)$ is as  follows:

 $$x=(\{u_1,u_2\}, \{t_1\}, \{r_1\}) \sqsubseteq y=(\{u_1,u_2\}, \{t_1\}, \{r_1, r_2\}) \Rightarrow$$ $$h(x))=(\{u_1,u_2\}, \{t_1, t_2\}, \{r_1\}) \not\sqsubseteq h(y)=(\{u_1,u_2\}, \{t_1\}, \{r_1, r_2\}).$$

\end{example}

\begin{figure}[ht!]
\begin{center}

\begin{tabular}{ccc}
\begin{tabular}{cccc}

& $t_1$ & $t_2$ &$t_3$\\
\cline{2-4}
$u_1$ &\multicolumn{1}{|c|}{$\times$} & \multicolumn{1}{|c|}{$\times$} &  \multicolumn{1}{|c|}{} \\
\cline{2-4}
$u_2$ &  \multicolumn{1}{|c|}{$\times$} &  \multicolumn{1}{|c|}{$\times$} &  \multicolumn{1}{|c|}{} \\
\cline{2-4}
$u_3$ &  \multicolumn{1}{|c|}{} &  \multicolumn{1}{|c|}{} &  \multicolumn{1}{|c|}{} \\
\cline{2-4}

& \multicolumn{3}{c}{$r_1$}\\

\end{tabular} & \quad\quad &

\begin{tabular}{cccc}

& $t_1$ & $t_2$ &$t_3$\\
\cline{2-4}
$u_1$ &\multicolumn{1}{|c|}{$\times$} & \multicolumn{1}{|c|}{} &  \multicolumn{1}{|c|}{} \\
\cline{2-4}
$u_2$ &  \multicolumn{1}{|c|}{$\times$} &  \multicolumn{1}{|c|}{} &  \multicolumn{1}{|c|}{} \\
\cline{2-4}
$u_3$ &  \multicolumn{1}{|c|}{} &  \multicolumn{1}{|c|}{} &  \multicolumn{1}{|c|}{} \\
\cline{2-4}

& \multicolumn{3}{c}{$r_2$}\\

\end{tabular}

\end{tabular}

\end{center}
\caption{A small example with Bibsonomy data}\label{counterEx1}
\end{figure}

\begin{definition} (\cite{Ganter:1999}, p.237, \cite{Ganter:2012})
  A relation $R\subseteq G\times M$ is called a Ferrers relation
  iff there are subsets   $A_1\subset A_2\subset A_3\ldots\subseteq G$
and $M\supseteq B_1\supset B_2\supset  B_3\supset\ldots$   such that
$R=\bigcup_{i}A_i\times B_i$.

 $R$ is called a Ferrers relation of concepts of $(G,M,I)$ iff
  there are formal concepts $(A_1,B_1)\le (A_2,B_2)\le (A_3,B_3)\le
  \ldots$ such that $R=\bigcup_{i}A_i\times B_i$.
\end{definition}


\begin{proposition}\label{prop:ferrers}\citep{Ganter:2012}
  Any Ferrers relation $R\subseteq I$ is contained in a Ferrers
  relation of concepts of $(G,M,I)$.
\end{proposition}
\begin{corollary} Let $\K=(G,M,B,I)$ be a formal tricontext, and $\K^{MB}_X=(M,B,I_X)$ such that $(m,b) \in I_X$ iff $(g,m,b) \in I \cap X \times M \times B$, and $I_X$ be Ferrers relation of concepts of $\K^{MB}_X$.
 Operator $h$ is not monotone for every pair of trisets $(X,Y,Z)$ and $(X,Y_i,Z_i)$  such that $Y \subseteq Y_i$, $Z\subseteq Z_j$, and $(Y_i,Z_i) \leq (Y_j,Z_j)$ are concepts of $\K^{MB}_X$.
\end{corollary}

\subsection{Inconsistency of MCCS closure}

\subsubsection{Lossy hyperedge encoding and phantom edges}

In case of $k$-partite graph encoding we can meet information loss in a form of new hyperedges. Below we provide this encoding from polyadic contexts to multi-relation databases with $n$ types of entities.

Let $\K=(K_1, \ldots, K_n, I)$ be a polyadic formal context, then $\D=(E=K_1 \sqcup \ldots \sqcup K_n,t,\R,R)$ be the corresponding multi-relation database, where $t$ maps entities from $E$ into their types from $1$ to $n$, $R=\{\{i,j\} \mid i,j \in \{1,\ldots, n\}, i\neq j\}$ and $\mathcal{R} = \bigcup_{\{i, j\} \in R}
\R_{\{i, j \}}$ for the binary relations $\mathcal{R}_{\{i, j\}} = \{\{e_i , e_j\}\mid  e_i \in K_i , e_j \in
K_j \mbox{ and } e_i , e_j \mbox{ are related by } I\}$.

\begin{figure}[ht!]
	\centering
		\includegraphics[width=0.5\textwidth]{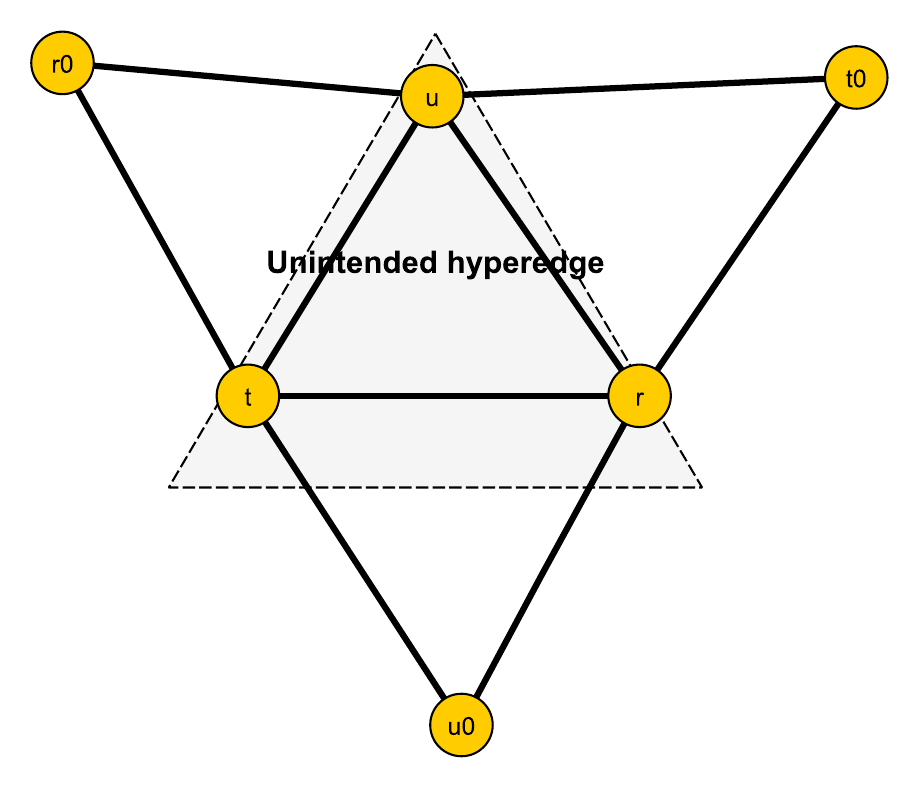}
	\caption{A phantom hyperedge as a structure loss}
	\label{fig:Loss}
\end{figure}

\begin{example}\label{ex:tri_loss}Imagine that we have three hyperedges $\{u,t,r_0\}, \{u,t_0,r\}, \{u_0,t,r\}$, and then encode them as edges in a 3-partite graph,
we obtain 
$$\{u,t\}, \{u,r_0\},  \{t,r_0\}, \{t_0,r\}, \{u,r\}, \{u,t_0\}, \{t,r\}, \{u_0,r\}, \mbox{and } \{u_0,t\}.$$

Since we now have $\{u,t\},  \{u,r\}, \mbox{and } \{t,r\}$ in our graph, we should inevitably decode a new hyperedge, $\{u,t,r\}$. See Figure \ref{fig:Loss}.
\end{example}

Taking the last fact into account and following the definition of MCCS or applying $g(\cdot)$ to respective CCS, we should obtain in general case a different or extra pattern(s) in addition to triconcepts in $k$-partite encoding. Thus in Example~\ref{ex:tri_loss} there are three MCCSs, $\{u, t_0,t,r\}$, $\{u,u_0,t,r\}$, and $\{u,t,r_0,r\}$, that are different from set representation of formal triconcepts, $\{u,t_0,r\}$, $\{u_0,t,r\}$, $\{u,t,r_0\}$, of the initial tricontext, respectively. 

\subsubsection{Closed but non-maximal patterns}

\begin{figure}[ht!]
	\centering
		\includegraphics[width=0.5\textwidth]{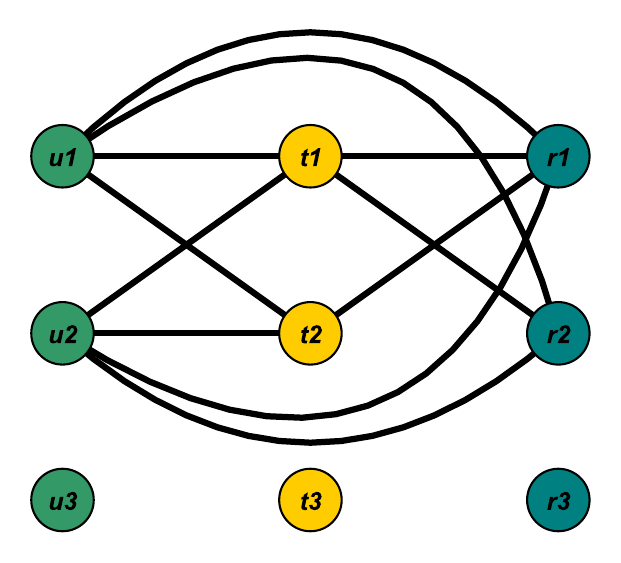}
	\caption{A counter example: closed but non-maximal patterns}
	\label{fig:LoicCounterEx}
\end{figure}

As one can see from the example in Table \ref{counterEx2}, the technical condition for idempotency of $g(\cdot)$ is fulfilled. The corresponding tripartite graph is depicted in Figure~\ref{fig:LoicCounterEx}.

\begin{table}[ht!]
\caption{A small example with Bibsonomy data}\label{counterEx2}
\begin{center}

\begin{tabular}{ccccc}
\begin{tabular}{cccc}

& $t_1$ & $t_2$ &$t_3$\\
\cline{2-4}
$u_1$ &\multicolumn{1}{|c|}{$\times$} & \multicolumn{1}{|c|}{$\times$} &  \multicolumn{1}{|c|}{} \\
\cline{2-4}
$u_2$ &  \multicolumn{1}{|c|}{$\times$} &  \multicolumn{1}{|c|}{$\times$} &  \multicolumn{1}{|c|}{} \\
\cline{2-4}
$u_3$ &  \multicolumn{1}{|c|}{} &  \multicolumn{1}{|c|}{} &  \multicolumn{1}{|c|}{} \\
\cline{2-4}

& \multicolumn{3}{c}{$r_1$}\\

\end{tabular} & \quad\quad &

\begin{tabular}{cccc}

& $t_1$ & $t_2$ &$t_3$\\
\cline{2-4}
$u_1$ &\multicolumn{1}{|c|}{$\times$} & \multicolumn{1}{|c|}{} &  \multicolumn{1}{|c|}{} \\
\cline{2-4}
$u_2$ &  \multicolumn{1}{|c|}{$\times$} &  \multicolumn{1}{|c|}{} &  \multicolumn{1}{|c|}{} \\
\cline{2-4}
$u_3$ &  \multicolumn{1}{|c|}{} &  \multicolumn{1}{|c|}{} &  \multicolumn{1}{|c|}{} \\
\cline{2-4}

& \multicolumn{3}{c}{$r_2$}\\

\end{tabular}

& \quad\quad &

\begin{tabular}{cccc}

& $t_1$ & $t_2$ &$t_3$\\
\cline{2-4}
$u_1$ &\multicolumn{1}{|c|}{} & \multicolumn{1}{|c|}{} &  \multicolumn{1}{|c|}{} \\
\cline{2-4}
$u_2$ &\multicolumn{1}{|c|}{} & \multicolumn{1}{|c|}{} &  \multicolumn{1}{|c|}{} \\
\cline{2-4}
$u_3$ & \multicolumn{1}{|c|}{} & \multicolumn{1}{|c|}{} &  \multicolumn{1}{|c|}{} \\
\cline{2-4}

& \multicolumn{3}{c}{$r_3$}\\

\end{tabular}

\end{tabular}
\end{center}

\end{table}

However for the CCS pattern $X=\{u_1,u_2, t_1, r_1\}$ the result of $g(X)$ coincides with $X$ but it is not maximal. Indeed, there exist two maximal closed and connected patterns corresponding to triconcepts,  $X \cup t_2 =\{u_1,u_2, t_1, t_2, r_1\}$ and  $X \cup r_2 =\{u_1,u_2, t_1, r_1, r_2\}$.

It is so, since $Comp(X)=X \cup \{t_2, r_2\}$, but $Comp(X \cup t_2) = X \cup t_2$ and $Comp(X \cup r_2)=X \cup r_2$.

\begin{proposition} Let $\F_\D$ be a CCS system and $\mathcal H \subseteq \F_\D$ such that $|\mathcal H|\geq 2$, every $H \in \mathcal H$ is maximal and there exists a CCS $X=\bigcap_{H \in \mathcal H} H \neq \emptyset$, then $g(X)=X$ but $X$ is not an MCCS.
\end{proposition}

\Proof{
Since there exist more than two MCCSs $H_i, H_j \in H$, we obtain $X\subset H_i$ and $X\subset H_j$. Therefore $H_i\setminus X \subseteq Comp(X)$ and
$H_j\setminus X \subseteq Comp(X)$. However for $h_i \in H_i$ and $h_j \in H_j$, $Comp(X \cup {h_i}) \neq Comp(X \cup {h_j})$ since otherwise it violates maximality of $H_i$. }

Let us introduce generalised Ferres relation of $n$-concepts (for 3-adic case see \citep{Glodeanu:2013}) .

\begin{definition}
  A relation $R\subseteq K_1 \times \cdots \times K_n$ is called a generalised Ferrers relation
  iff $\exists j \in \{1,\ldots,n\} \forall i \in \{1,\ldots,n\}\setminus \{j\}$ $A_{1i}\subset A_{2i}\subset A_{3i}\ldots\subseteq K_i$, and $K_j \supseteq A_{1j}\supset A_{2j}\supset  A_{3j}\supset\ldots$ such that 
$R=\bigcup_{k} A_{k1}\times \ldots \times A_{kn}$.

 $R$ is called a Ferrers relation of n-concepts of $(K_1, \ldots,K_n,I)$ iff
  there are formal n-concepts $(A_{11}, \ldots, A_{1n}) \lesssim_k (A_{21},\ldots,A_{2n}) \lesssim_k (A_{31},\ldots,A_{3n}) \lesssim_k
  \ldots$ such that $R=\bigcup_{i} A_{i1}\times \ldots \times A_{in}$.
\end{definition}

\begin{corollary}
Let $\K=(K_1, \ldots, K_n, I)$ be a polyadic formal context such that $I$ is a Ferrers relation of $n$-concepts and $\D=(E=K_1 \sqcup \ldots \sqcup K_n,t,\R,R)$ be the corresponding multi-relation database. Operator $g$ does not produce an MCCS for $flat(A_{j1} \cap A_{i1},  \ldots, A_{jn} \cap A_{in})$ obtained from any  pair of concepts of $\K$, $(A_{i1},  \ldots, A_{in}) \lesssim_k (A_{j1},  \ldots, A_{jn})$, where $A_is\neq A_js$, $s, k \in \{1,\ldots,n\}$ and $s\neq k$.
\end{corollary}

\section{Closure operator for triconcepts}\label{sec:triclo}

There are $n$-contexts, where $h(\cdot)$ is not a closure that results in formal concepts because of non-idempotency and closure operator $g(\cdot)$ produces CCSs that are not necessary maximal, e.g. caused by non-uniqueness of possible extensions of input patterns. Moreover, the lossy data encoding by $n$-partite graph instead of $n$-partite hypergraph results in phantom $n$-adic edges and extra elements in resulting patterns. 

So, to overcome the difficulty at least for generation of $n$-concepts we may adjust the set systems such that $h(\cdot)$ and $g(\cdot)$ could operate. Informally, we need to weed all patterns or phantom hyperedges that result in undesirable behaviour of $h(\cdot)$ and $g(\cdot)$, the candidates to closure operators.

\begin{definition}
Let $\K=(K_1, K_2, K_3, I)$ be a triadic formal context. A triset $S$ is called a (maximal) switching generator of the context $\K$ iff $S=tuple(flat(c_1) \cap flat(c_2))\neq \emptyset$, where $c_1$ and $c_2$ are concepts of $\K$.
\end{definition}

\begin{theorem}\label{thrm:weeding}
Let $\K=(K_1, K_2, K_3, I)$ be a triadic formal context. The set system $\F_{\K\ominus\mathcal{S}}=\mathcal T\setminus\mathcal{S}$ is a correct set system for formal triconcept generation by $h(\cdot)$ in $\K$,  where $\mathcal{T} = \{ (X,Y,Z)  \in  2^G\times 2^M\times 2^B \mid (X,Y,Z) \subseteq I \}$ and $\mathcal{S}=\{S \mid  S \mbox{ is a switching generator of } \K\}$.
\end{theorem}

\Proof{
Since there is no a switching generator in $\F_{\K\ominus\mathcal{S}}$, monotony of $h(\cdot)$ is fulfilled. 

Assume that monotony is violated by trisets $x$ and $y$, that is $x \sqsubseteq y  \to h(x)\not\sqsubseteq h(y)$. By extensivity of $h(\cdot)$ and transitivity of $\sqsubseteq$, it implies $x \sqsubseteq h(x)$ and $x \sqsubseteq h(y)$. Hence, $x \sqsubseteq  tuple(flat(h(x)) \cap flat(y))$, i.e. $x$ is a switching generator. Contradiction.

Since every formal triconcept is not a switching generator, none of triconcepts has been deleted from $\F_{\K\ominus\mathcal{S}}$. 

}

As for phantom triadic edges, unfortunately it is not possible to delete them from $\R$ since each phantom triadic edge $\{e_i, e_j, e_k\}$ is composed by $\{e_i,e_j\}$, $\{e_j, e_k\}$, and $\{e_k,e_i\}$, which are parts of ``real'' triadic hyperedges.

Let $\K$ be a formal tricontext and $\D$  be the corresponding multi-relational database, $\mathcal P=\{tuple(e) \mid  e=\{e_i,e_j,e_k\} \mbox{ is a phantom edge in } \R \}$ then a test whether an MCCS forms triset can be done as follows:

\begin{enumerate}

\item For an MCCS $s$ form $tuple(s)=(X,Y,Z)$;
\item Check whether $t=X \times Y \times Z \setminus e$ forms a triset of $\K$, where $e \in \mathcal P$;
\item If yes, then output $t$;
\item Delete $s$ from the output otherwise.

\end{enumerate}  

To make sure that $t$ is a triconcept, one need to check $h(t)=t$.

Since traditionally closure operators were introduced for partial orders over set inclusion, we would like to avoid dealing with preoder $\sqsubseteq$ over trisets and work with set inclusion of their set representations instead.

For tricontext $\K=(K_1,K_2,K_3, I)$ we consider a family of operators 
\begin{multline*}
\{\sigma_{ijk}| \sigma_{ijk}: 2^{K_1} \times 2^{K_2} \times 2^{K_3} \to 
2^{K_1} \times 2^{K_2} \times 2^{K_3} \mbox{ such that } \\
 \sigma_{ijk}: (X_1,X_2,X_3) \mapsto (Y_1,Y_2,Y_3), \mbox{ where }\\
Y_i = (X_j \times X_k)^{(i)}, Y_j= (Y_i \times X_k)^{(j)}, Y_k=(Y_i \times Y_j)^{(k)} \mbox{, where } \{i,j,k\}=\{1,2,3\}\}.
\end{multline*}

The cardinality of the family is $3!=6$ and $n!$ for its n-ary case generalisation.

\begin{proposition}\label{prop:nocomm} Operators $\sigma_{ijk}(\cdot)$ are not commutative, i.e. $\sigma_{ijk}(\sigma_{lmn}(\cdot))\neq \sigma_{lmn}(\sigma_{ijk}(\cdot))$, where $(i,j,k)\neq (l,m,n)$ and $\{i,j,k\}=\{1,2,3\}$.
\end{proposition}

\Proof{ Consider a tricontext given below.

\begin{tabular}{ccccc}
\tabcolsep=0.1cm
\begin{tabular}{ccccc}

& $t_1$ & $t_2$ & $t_3$ & $t_4$ \\
\cline{2-5}
$u_1$ &\multicolumn{1}{|c|}{$\times$} & \multicolumn{1}{|c|}{$\times$}  &\multicolumn{1}{|c|}{$\times$}  &\multicolumn{1}{|c|}{$\times$} \\
\cline{2-5}
$u_2$ &  \multicolumn{1}{|c|}{} &  \multicolumn{1}{|c|}{$\times$}  &\multicolumn{1}{|c|}{$\times$}  &\multicolumn{1}{|c|}{$\times$}  \\
\cline{2-5}
$u_3$ &  \multicolumn{1}{|c|}{} &  \multicolumn{1}{|c|}{$\times$}  &\multicolumn{1}{|c|}{$\times$}  &\multicolumn{1}{|c|}{$\times$}  \\
\cline{2-5}
$u_4$ &  \multicolumn{1}{|c|}{} &  \multicolumn{1}{|c|}{}  &\multicolumn{1}{|c|}{}  &\multicolumn{1}{|c|}{$\times$}  \\
\cline{2-5}

& \multicolumn{4}{c}{$r_1$}\\

\end{tabular} &  &
\tabcolsep=0.1cm
\begin{tabular}{ccccc}

& $t_1$ & $t_2$ & $t_3$ & $t_4$ \\
\cline{2-5}
$u_1$ &\multicolumn{1}{|c|}{} & \multicolumn{1}{|c|}{$\times$}  &\multicolumn{1}{|c|}{$\times$}  &\multicolumn{1}{|c|}{$\times$} \\
\cline{2-5}
$u_2$ &  \multicolumn{1}{|c|}{} &  \multicolumn{1}{|c|}{$\times$}  &\multicolumn{1}{|c|}{$\times$}  &\multicolumn{1}{|c|}{$\times$}  \\
\cline{2-5}
$u_3$ &  \multicolumn{1}{|c|}{} &  \multicolumn{1}{|c|}{}  &\multicolumn{1}{|c|}{$\times$}  &\multicolumn{1}{|c|}{$\times$}  \\
\cline{2-5}
$u_4$ &  \multicolumn{1}{|c|}{} &  \multicolumn{1}{|c|}{}  &\multicolumn{1}{|c|}{}  &\multicolumn{1}{|c|}{}  \\
\cline{2-5}

& \multicolumn{4}{c}{$r_2$}\\

\end{tabular} &  &

\tabcolsep=0.1cm
\begin{tabular}{ccccc}

& $t_1$ & $t_2$ & $t_3$ & $t_4$ \\
\cline{2-5}
$u_1$ &\multicolumn{1}{|c|}{} & \multicolumn{1}{|c|}{}  &\multicolumn{1}{|c|}{}  &\multicolumn{1}{|c|}{$\times$} \\
\cline{2-5}
$u_2$ &  \multicolumn{1}{|c|}{} &  \multicolumn{1}{|c|}{}  &\multicolumn{1}{|c|}{}  &\multicolumn{1}{|c|}{}  \\
\cline{2-5}
$u_3$ &  \multicolumn{1}{|c|}{} &  \multicolumn{1}{|c|}{}  &\multicolumn{1}{|c|}{}  &\multicolumn{1}{|c|}{}  \\
\cline{2-5}
$u_4$ &  \multicolumn{1}{|c|}{} &  \multicolumn{1}{|c|}{}  &\multicolumn{1}{|c|}{}  &\multicolumn{1}{|c|}{}  \\
\cline{2-5}

& \multicolumn{4}{c}{$r_3$}\\

\end{tabular}

\end{tabular}

The system of all switching generators $\mathcal{S}$ contains $s_1=\{u_1,t_4,r_1\}$ and $s_2=\{u_1,u_2, t_3,t_4, r_1\}$.

$s_1$ proves that $\sigma_{i\_\_}(\cdot)\neq\sigma_{j\_\_}(\cdot)\neq\sigma_{k\_\_}(\cdot)$ and

$s_2$ proves that $\sigma_{ijk}(\cdot)\neq\sigma_{ikj}(\cdot)$ for $\{i,j,k\}=\{1,2,3\}$.

The fact that $\sigma_{lmn}\sigma_{ijk}(\cdot)=\sigma_{ijk}(\cdot)$ proves the proposition.
}

\begin{theorem}\label{thrm:noclos}  For $\K=(K_1,K_2,K_3,I)$ and the associated triset system  $\mathcal T$ there is no an associated closure operator in case there exist at least two concepts  $c_1=(X_1,Y_1,Z_1)$ and $c_2=(X_1,Y_2,Z_2)$ such that they have the common non-empty maximal switching generator $s$, i.e. $tuple(flat(c_1) \cap flat(c_2)) \neq \emptyset$.
\end{theorem}

\Proof{ Let $\sigma$ be a closure operator for $\K$.
Since $s \sqsubset c_1$ and $s \sqsubset c_2$ then $\sigma(s)$ should result in $c_i$ which is either $c_1$ or $c_2$ (or one of other concepts $c_k$ with $s \sqsubset c_k$ if any exist). So let $\sigma(s)=c_i$ and consider $s \sqsubseteq c_j$; it implies that $\sigma(s)=c_i \not \sqsubseteq \sigma(c_j)=c_j$ for $i\neq j$, and $\{i,j\}=\{1,2\}$. Contradiction.
}

As it has been shown, $\F_{\K\ominus S}$ is a correct set system for $h(\cdot)=\sigma_{123}(\cdot)$ being a closure operator. It is easy to see that this system is correct for $\sigma_{ijk}(\cdot)$.

To summarise properties of $\F_{\K\ominus S}$ and show its difference from set systems in \citep{Boley:2010,Spyropoulou:2014} we recall the following properties of set systems.

\begin{definition}
A non-empty set system $(E,\F)$ is called 

1. accessible if for all $X \in \F \setminus \{\emptyset\}$ there is an $e \in X$ such that $X \setminus \{e\} \in \F$,

2. an independence system if $Y \in F$ and $X \subseteq Y$ together imply $X \in F$ ,

3. confluent if for all $I, X, Y \in  \F$ with $\emptyset \in I \subseteq X$ and $I \subseteq Y$ it holds
that $X \cup Y \in F$.

4. strongly accessible if it is accessible and for all $X, Y \in F$ with
$X \subset Y$, there is an $e \in Y \setminus X$ such that $X \cup \{e\} \in \F$ .

\end{definition}

\begin{proposition} \label{prop:propert}
1) Set system $\F_{\K\ominus S}$ of all sets that form trisets is accessible and 2) not independent. 3) It is not a closure system. 4) It is confluent. 5) It is strongly accessible.
\end{proposition}

\Proof{
1. Every set of $\F_{\K\ominus S}$ forms a triset $t$. Even if it contains some switching generator $s$, we can  then remove any $e \in flat(s)$ from $t$, the resulting set $flat(t) \setminus {e}$ is in $\F_{\K\ominus S}$ (switching generator free system) since it is a triset and contains at least one element not included in a switching generator. Empty set (or empty set of triples) is not in $\F_{\K\ominus S}$ because it is a universal switching generator.

2. Since some concepts may contain switching generators by triset set inclusion, it implies that these switching generators are not $\F_{\K\ominus S}$.

3. On the contrary, every pair of concepts $X,Y \in \F_{\K\ominus S}$ implies that $X \cap Y \notin \F$  (anti-sharing).

4. Since there is no such non-empty $I \in \F_{\K\ominus S}$ being a triset of two different concepts it trivially holds. 

5.  If $X \subset Y$ for $X,Y \in \F_{\K\ominus S}$, $X$ does not form a formal concept (because of antiordinal relations) or switching generator. So adding any element $e$ from $Y\setminus X$ leaves $X \cup \{e\}$ being a triset.
}

A detailed study of algorithmic issues is out of scope the paper,  however, \cite{Boley:2010} reported a simple algorithm for Problem~\ref{prbl:1}, i.e. listing of all fixed points of a partially defined closure operator, which is correct for strongly accessible set systems.

\begin{problem} \label{prbl:1} (list-closed-sets) Given a set system $(E, \F)$ with $\emptyset \in \F$ and a closure operator $\sigma:  \F \to \F$ , list the elements of $\sigma(\F)=\{F\mid  F \in \F: \sigma(F)=F \}$.
\end{problem}


It is questionable whether the weeding step can be efficiently incorporated into closure listing algorithm. Thus, in the worst case, i.e. for power tricontext $\K=(\{1\ldots k\}, \{1\ldots k\}, \{1\ldots k\}, \neq)$,  the number of triconcepts equals $3^k$ \citep{Biedermann:1998}. The number of switching generators is  greater than that of the concepts of $\K$ for $k>2$ and not polynomial as given in~Theorem~\ref{thrm:numsw}

\begin{theorem}\label{thrm:numsw} For a power tricontext $\K=(\{1\ldots k\}, \{1\ldots k\}, \{1\ldots k\}, \neq)$  the number of switching generators is $4^k-3^k$.
\end{theorem}

\Proof{
One can prove the theorem  by direct calculation of the triple sum below:

$$\sum\limits_{k_1=0}^{k-1}\sum\limits_{k_2=0}^{k-k_1-1}\sum\limits_{k_3=0}^{k-k_1-k_2-1}C^{k_1}_kC^{k_2}_{k-k_1}C^{k_3}_{k-k_1-k_2}.$$
}

Theorems \ref{thrm:weeding},\ref{thrm:noclos} and propositions \ref{prop:propert},\ref{prop:nocomm} can be generalised for $n$-ary case in a similar way.  For example, the general version of Theorem~\ref{thrm:numsw} is provided as Theorem~\ref{thrm:numswn}.

\begin{theorem}\label{thrm:numswn} For a power polyadic $n$-context $\K=(\{1\ldots k\}, \ldots, \{1\ldots k\}, \neq)$  the number of switching generators is $(n+1)^k-n^k$.
\end{theorem}

In \cite{Kuznetsov:2004}, the complexity of the problem ``Number of all concepts'' was addressed. Thus this problem is $\#P$-complete. Theorem~\ref{thrm:sharpp} provides a similar justification, showing that the problem ``Number of all (maximal) switching generators'' is intractable. To avoid complex technicalities we prove the theorem for $n=2$.

\begin{theorem}\label{thrm:sharpp} 
The following problem ``Number of all (maximal) switching generators'' is $\#P$-complete:

 Input: Context $\mathbb{K}=(G,M,I)$

 Output: The number of all (maximal) switching generators of the context $\mathbb{K}$, i.e. $|\mathcal{S}|
 $.
\end{theorem} 
\Proof{ We reduce the following $\#P$-complete problem to ours: ``The number of binary vectors that satisfy monotone 2-CNF of the form $C=\bigwedge_{i=1}^{s} ( x _{i,1} \vee x_{i,2})$'':

\emph{Input}: Monotone (without negation) CNF with two variables in each   disjunction $C=\bigwedge_{i=1}^{s} ( x _{i,1} \vee x_{i,2}),$ where $x_{i,1},x_{i,2} \in X=\{x_1, \ldots, x_k\}$ for all $i=\overline{1,s}$. 

\emph{Output}: Number of binary $k$-vectors (corresponding to the values of variables) that satisfy CNF $C$.

First, we construct 2-DNF $D$, the negation of $C$: $\bigvee_{i=1}^{s} ( \overline{x} _{i,1} \wedge \overline{x}_{i,2}),$. Each conjunction is denoted  $D_i=( \overline{x} _{i,1} \wedge \overline{x}_{i,2})$, $i=\overline{1,s}$. The set of binary vectors that satisfy $D$ is a union of the sets of binary vectors that satisfy a certain $D_i$. Each disjunction is satisfied by every binary $k$-vector with $k-2$ ones and two zeros in $i_1$-th and $i_2$-th components.

We reduce this problem to that of the number of switching generators by constructing the following context $\mathbb{K}=(G,M,I)$. The set of attributes is $M=\{m_1,\dots,m_k\}\cup \bigcup_{i=1}^s \{ m^{i,k-1}, m^{i,k} \}$, where elements of $\tilde{M}=\{m_1,\dots,m_k\}$ are in one-to-one correspondence with variables from $X$. For some conjunction $D_i, i=\overline{1,s}$, we construct a context $\mathbb{K}_i=(G_i,M_i,I_i)$, where the set of attributes is $M_i=\tilde{M}\setminus \{m_{i,1},m_{i,2}\} \cup \{m^{i,k-1}, m^{i,k}\}:=\{m^{i,1}, \ldots, m^{i,k}\}$, the set of objects is $G_i=\{g_i^0,g_i^1,\ldots, g_i^{k-2},g_i^{k-1}, \ldots, g_i^{2k-2}\}$, and the relation $I_i \subseteq M_i\times G_i$ is defined as follows: $\{g_i^0\}'=M_i\setminus \{m^{i,k}\}$, $\{g_i^j\}'=M_i\setminus \{m^{i,j}, m^{i,k}\}$ for $j=\overline{1,k-2}$, $\{g_i^j\}'=M_i\setminus \{m^{i,j},m^{i,k-1},m^{i,k} \}$ for  $j=\overline{k-1,2k-4}$, $\{g_i^{k-3}\}'=M_i\setminus \{m^{i,k-1},m^{i,k} \}$, and $\{g_i^{2k-2}\}'= \{m^{i,k-1},m^{i,k}\}$. The context $\mathbb{K}$ is constructed as follows:  $\K=(\bigsqcup_{i=1}^{s} G_i, M, \bigsqcup_{i=1}^{s} I_i)$ with only one $\{g_i^{2k-2}\} \in G_i$ (the remaining $G_j$-s do not contain  $\{g_j^{2k-2}\}$).

First, we show that every switching generator of $\mathbb{K}$ corresponds to a $k$-vector that satisfies $D$. Every switching generator of $\mathbb{K}$ is a switching generator of $\mathbb{K}_i$ for some $i$, which can be not unique.  It is easy to see that intents of the context $\mathbb{K}_i$ form the power set of $M_i$, denoted by $2^{M_i}$. Elements of $2^{\tilde{M}}$ are in one-to-one correspondence with binary $n$-vectors, where components are in one-to-one correspondence with elements of $M$ with the same number. Since for every non-empty $S \subseteq \tilde{M}\setminus \{m_{i,1},m_{i,2}\}$ there are concepts $(S',S)$ and $((S\cup m^{i,k-1})',S\cup m^{i,k-1})$, their switching generator is $((S\cup m^{i,k-1})',S)$. A vector of this form satisfies $D_i$, since it has zeros at $i_1$-th and $i_2$-th places. Therefore, this vector satisfies $D$. To prove that the switching generator is provided when $S=\emptyset$, one may check that $(\{g_i^{2k-2}\},\{m^{i,k-1}\})$ is the switching generator of  concepts $(\{g_i^{2k-2}\},\{m^{i,k-1},m^{i,k}\})$  and $(\{m^{i,k-1}\}', \{m^{i,k-1}\})$.

It remains to show that binary $k$-vectors that satisfy $D$ are in one-to-one correspondence with intents of $\mathbb{K}$. In fact, each binary $k$-vector $v$ that satisfies $D$, satisfies $D_i$ for some $i$ (this $i$ may be not unique). Then this vector has zero $i_1$-th and $i_2$-th components. Therefore, the corresponding set of attributes $A$ belongs to $\tilde{M} \setminus \{m_{i,1},m_{i,2}\} \subseteq 2^{M_i}$. If  $A\neq\emptyset$, then concepts $(A',A)$ and $((A\cup m^{i,k-1})',A\cup m^{i,k-1})$ has switching generator is $((A\cup m^{i,k-1})',A)$. If  $A=\emptyset$, then there is a unique switching generator $(\{g_i^{2k-2}\},\{m^{i,k-1}\})$ for some $G_i$.

The one-to-one correspondence between the switching generators of concepts of $\mathbb{K}$ and binary $k$-vectors satisfying $D$ is realised. Thus, if we know the number of all switching generators of concepts of $\mathbb{K}$, we obtain the number of all vectors satisfying $D$ and, therefore, that of the vectors satisfying $C$. The reduction is polynomial in the input size, since the context $\mathbb{K}$ has $|M|=k+2s$ attributes and $|G|=s(2k-2)+1$ objects.
}

Similar theorem can be proved for $n=3$.

\begin{corollary}\label{croll:sharpp} 

The problem ``Number of all (maximal) switching generators for $n=3$'' is $\#P$-complete:

 Input: Context $\mathbb{K}=(G,M,B,Y)$

 Output: The number of all (maximal) switching generators of the context $\mathbb{K}$, i.e. $|\mathcal{S}|
 $.

\end{corollary}

The proof can be done in a similar way to the dyadic case, where $\mathbb{K}_i=(G_i,M,B_i,Y_i)$ for each conjunction $D_i$ should have $B_i=\{b_i^1, b_i^2\}$ (in fact, it plays a role of $\{m^{i,k-1}, m^{i,k}\}$ from Theorem~\ref{thrm:sharpp}) such that each $A$ will result in two triconcepts $(U,A,\{b_i^1, b_i^2\})$ and  $(V,A,\{b_i^1\})$, $U \subseteq V$ with their maximal switching generator $(U,A,\{b_i^1\})$.

\section{Related work}\label{sec:relw}

In fact, one of the first methods for triconept enumeration was \textsc{TRIPAT} \citep{Ganter:1994} adopted Ganter's \textsc{Next Closure} algorithm in a nested manner for two-adic contexts generated from an input tricontext; it had been done even before the first formal treatment of 3FCA by \cite{Lehmann:1995}. This idea has been incarnated later in \textsc{TRIAS} for triconcepts enumeration with component-wise size constraints \citep{Jaschke:2006}.

Due to intrinsic complexity of exhaustive enumeration of triconcepts and closed $n$-sets, the research focus has shifted to constrained pattern mining and searching for different relaxations. Thus, after the release of \textsc{DataPeeler} \citep{Cerf:2009}, the \textsc{Fenster} algorithm for faul-tolerant pattern discovery has been proposed \citep{Cerf:2013}; the latter includes closed $n$-set mining as a particular case, allowing not all tuples inside dense $n$-sets to be present. Another approach, is the so-called OAC-triclustering for mining dense trisets \citep{Ignatov:2015} results in no more patterns than the number of tuples in an input relation having a fruitful property of containment of all triconcepts for a given tricontext within the resulting collection of trisets w.r.t. to component-wise inclusion under a properly selected minimal density constraint. 
A different approximation of triconcept can be realised within least square error minimisation criterion (see \textsc{TriBox},  \cite{Mirkin:2011}), which lead to a density-based pattern quality measure, namely the squared density of a triset (in sense of \citep{Cerf:2009}) multiplied by its size, thus, expressing trade-off between the high number of non-missing tuples inside and the large size.

One more direction is to use factorisation to select only a(n) (optimal) subset of triconcepts, which are factors to decompose an input three-way Boolean tensor \citep{Glodeanu:2013,Belohlavek:2013}. Closed sets are helpful for mining numeric contexts as well; thus, \cite{Kaytoue:2013} used 3FCA for searching maximal inclusion biclusters of constant values by treatment of attribute values as conditions. \cite{Spyropoulou:2014} proposed MCCS patterns and the associated closure operator for $n$-partite graphs working with multi-relational data. They also performed experimental comparison their \textsc{RMiner} with \textsc{DataPeeler}, which is not fully correct since $n$-ry relations being encoded as $n$-partite graphs result in phantom edges. Note that, in FCA domain, there is Relational Concept Analysis devoted to treatment of multi-relational data \citep{Hacene:2013}. The group that works on MCCSs has recently proposed Complete Connected Proper Subsets (CCPS) to deal with relational data with structured attributes \citep{Lijffijt:2016}, i.e. attributes with ordered values like real numbers, geographic location, time intervals, etc. Note that in FCA domain, to deal with data of complex description the so called Pattern Structures were proposed more than decade ago by \cite{Ganter:2001} and found many succesfull applications \citep{Kaytoue:2015}.

There is an interesting connection between biclique operators, their associated graphs \citep{Crespelle:2015}, and switching generators;  in these graphs, two vertices (maximal biclques) are connected if they have a non-empty intersection, which, under some conditions, can be the switching generator of those biclques, i.e. concepts.

\section{Conclusion}\label{sec:concl}

 The recent candidates to be closure operators related to triconcepts  are not always consistent with either the definition of closure operator or triconcept ($n$-concept). We considered partially defined closure operators for triconcept generation that solve the problem.  It is easy to obtain their $n$-adic versions and generalise current results. However the open question at the moment is whether recent closure-based algorithms for pattern mining reported in the relevant literature may benefit from this new bit of knowledge. Even though their basic definitions can be refined to fulfill necessary requirements, as we have seen, it might be costly or even intractable. Thus, an interesting prospective result could be a polynomial time check whether the current context is switching generators free (excluding $\emptyset$) or has a polynomial number of switching generators; one of the switching generators free examples is $\K=(\{1\ldots m\}, \ldots, \{1\ldots m\}, =)$.

\subsubsection*{Acknowledgments} The author would like to thank Lo\"{i}c Cerf, Eirini Spyropoulou, Boris Schminke, Dmitry Gnatyshak, Sergei Kuznetsov, Sergei Obiedkov, Bernhard Ganter, Jean-Francois Boulicaut, Mehdi Kaytoue, Amedeo Napoli, Boris Mirkin, Lhouri Nourine, Engelbert Mephu Nguifo and Jaume Baixeries. The study was implemented in the framework of the  Basic Research Program at the National Research University Higher School of Economics in 2015--2017 and in the Laboratory of Intelligent Systems and Structural Analysis. The author was also partially supported by Russian Foundation for Basic Research.













\section*{References}

\bibliography{triclo}

\end{document}